# The velocity dispersion profiles of clusters of galaxies: a cosmological test and the sampling effect

Y. P. Jing and G. Börner
*Max-Planck-Institut für Astrophysik, Karl-Schwarzschild-Strasse 1, 85740 Garching, Germany*
*E-mail: jing@mpa-garching.mpg.de; grb@mpa-garching.mpg.de*

## ABSTRACT

In this paper we investigate the velocity dispersion profiles of clusters of galaxies for seven cosmological models. One model is the SCDM model, and the others are six low-density models with the density parameter $\Omega = 0.1$, 0.2 or 0.3 and with or without a cosmological constant $\Lambda = 1 - \Omega$. We find that the velocity dispersion profiles depend both on $\Omega$ and on $\Lambda$. For $\Lambda = 0$, the profiles are steeper in a lower-$\Omega$ model than in a higher-$\Omega$ one. The cosmological constant significantly weakens the dependence on $\Omega$: the difference in the profile distributions between two flat models is much smaller than that between the two corresponding open models with the same $\Omega$. These results in principle can be used to constrain the cosmological parameters when a large sample of the velocity dispersion profiles is available.

Motivated by the practical situation that a sample of $\sim 100$ clusters with $\sim 100$ measured redshifts per cluster is still the best sample available in the foreseen future, we examine carefully to what degree the cosmological parameters can be constrained with the velocity dispersion profiles of such a sample of clusters. The limited sampling around clusters and the limited number of clusters seriously degrade the discriminative power of the velocity dispersion profiles among cosmological models. We find that the five models of $\Omega \geq 0.2$ cannot be distinguished by this type of observation. Due to the limited sampling, one should be very cautious in extracting information about the density profile and/or the dynamics around single clusters from the diluted velocity dispersion profiles

**Key words:** galaxies: clustering - large-scale structure of the universe - cosmology: observations - dark matter

## 1 INTRODUCTION

Clusters of galaxies are the largest gravitationally bound objects which have collapsed very recently. The study of these objects can provide important clues to our understanding of large-scale structure formation and to the existence of dark matter in the universe. The velocity dispersion profile $\sigma_v(r_p)$, i.e. the velocity dispersion as a function of projected distance $r_p$ from the cluster center, is one of the important quantities which are used to describe properties of clusters. Not only is $\sigma_v(r_p)$ closely related to the density profile $n(r)$ of the cluster through dynamical equations, but it also is an indicator of the dynamical state of the cluster. Therefore, the velocity dispersion profiles would provide, at least in principle, a very useful test for cosmological models.

In this paper, we will examine how the velocity dispersion profiles depend on cosmological models. We will focus on their shapes, since the distribution function of their amplitudes (i.e. the velocity dispersion function) has already been a subject of many earlier studies (Frenk et al. 1990; Jing & Fang 1994, Zabludoff & Geller 1994; Crone & Geller 1995; Henry & Arnaud 1992; Bartlett & Silk 1993). Recent works by Crone, Evrard & Richstone (1994) and Jing et al. (1995; hereafter JMBF95) have shown that the shape of density profiles of clusters depends on cosmological parametres: clusters in an Einstein-de-Sitter universe have much flatter density profiles than those in an open universe of the density parameter $\Omega \leq 0.3$. The cosmological constant $\Lambda$ also sensitively influences the shape of density profiles, i.e. the density profiles of clusters in a low-density flat universe ($\Omega + \Lambda = 1$) are flatter than in a corresponding open universe with the same $\Omega$. Since the density profiles and the velocity dispersion profiles are closely related, one would expect that the shape of the velocity dispersion profiles also sensitively depends on the cosmological parameters. In this work we shall quantify this dependence.

We shall use the $P^3M$ N-body simulations of JMBF95 to study this problem. These simulations have been used to study the substructures and density profiles of clusters in cosmological models (JMBF95). The simulations, designed to study properties of clusters in different models, cover a large space of cosmological parameters (a total of 7 models). Each rich cluster consists of $\sim 1000$ particles, so the density and velocity fields around clusters are well sampled. Therefore these simulations are very suitable for our purposes here.

Clusters in the real universe, however, can not be sampled so densely by observations as in our simulations. Even with present observational facilities, it is still not an easy task to measure redshifts of $\sim 100$ members per cluster for a sample of $\sim 100$ clusters. The ESO Key-Programme on Nearby Galaxy Clusters is attempting to accomplish such a task (Giuricin et al. 1994). For a practical application of the velocity dispersion profiles to tests of cosmological models, a crucial question is *whether* $\sim 100$ *cluster members are sufficient to faithfully map the velocity fields around clusters or whether* $\sim 100$ *members can give an unbiased estimate of the cluster velocity dispersion profile.* In this work, we shall examine this sampling effect in detail, to study how the sampling biases the determination of the individual velocity dispersion profiles as well as their statistical distributions. From this study we will know to what extent the cosmological parameters can be constrained by the shape of the velocity dispersion profiles, if the cluster sample contains some 100 clusters with $\sim 100$ measured redshifts per cluster.

## 2 A BRIEF DESCRIPTION OF N-BODY SIMULATIONS AND CLUSTER IDENTIFICATION

The simulations used here are the $P^3M$ N-body simulations of JMBF95 for seven cosmological models. Of these models, one is the standard CDM model, and the others are six low-density CDM models of which three are flat with $\Lambda = 1 - \Omega$ and the other three are open with $\Lambda = 0$. The density parameters $\Omega$ of the low-density models are either 0.1, 0.2 or 0.3. We will use the same notations for these models as JMBF95, i.e. SCDM for the Standard CDM model, FL0$i$ and OP0$i$ ($i = 1, 2, 3$) for the FLat models and the OPen models with $\Omega = 0.1$, 0.2, 0.3 respectively. For example, FL03 is the flat model with $\Omega = 0.3$ and OP01 is the open model with $\Omega = 0.1$. The simulation box is a comoving cubic volume of $128^3\,h^{-3}\mathrm{Mpc}^3$ for every model. $100^3$ or $128^3$ particles are used for the SCDM model, and $64^3$ particles for the low-density models. The effective force resolution is $0.1\,h^{-1}$Mpc comoving. Three realizations (one using $128^3$ particles) are run for the SCDM model, and five for each of the other models. For more details, we refer the readers to JMBF95.

Clusters in these simulations have been identified by JMBF95 based on the procedures proposed by Jing & Fang (1994; hereafter JF94). These clusters will be used for our following analysis. The identification procedures of JF94 consist of two steps. The first step is to find a group list based on the *friends-of-friends* algorithm (e.g., White et al. 1987; Jing et al. 1993). The identified groups may have very irregular shapes, and thus may have their centers not located in dense regions. The second step is then used, to search for gravitational potential minima around the groups and to identify the sites of the minima as cluster centers. The latter procedure guarantees that the cluster center is always at the densest spot of the particle distribution in the cluster region. The accuracy of the center positions thus determined is estimated to be $\sim 0.035\,h^{-1}$Mpc on a two-dimensional surface (JMBF95). An accurate measurement of center positions is essential for analyzing the velocity dispersion profile, since the velocity dispersion profile is expected to depend on the cluster center.

## 3 THE VELOCITY DISPERSION PROFILE

To calculate the velocity dispersion profile for a cluster, first we need to define cluster members. In this work, cluster members are defined as the dark matter particles within the turn-around radius $r_{ta}$ of the cluster center. With a spherical infall model, it is found that the ratio $\mathcal{R}_{ta} \equiv \mathcal{R}(r_{ta})$ of the mean mass density within $r_{ta}$ to the background density is approximately $5.6\Omega^{-0.6}$ (Peebles 1984; White, Efstathiou & Frenk 1993). We determine $r_{ta}$ by calculating the ratio $\mathcal{R}(r)$ within different radii $r$ around clusters. If the difference between $\mathcal{R}(r)$ and $\mathcal{R}_{ta}$ is less than 5%, the radius $r$ is regarded as the turn-around radius $r_{ta}$. Only clusters with more than 500 members will be analyzed. There are some 170 clusters in each of the low-density models and 150 clusters in the SCDM model.

Choosing one direction, say $\boldsymbol{x}_3$, as the line-of-sight, we transform the coordinates of cluster members from real space to redshift space. Then we calculate the velocity dispersion $\sigma_v(r_p)$ for cluster members with projected distances in the range $r_p \pm \Delta r_p/2$ from the cluster center. Here we choose three bins of fixed radii 0, 0.2, 0.5, and 1.0 $h^{-1}$Mpc. We calculate $\sigma_v$ for each direction of $\boldsymbol{x}_1$, $\boldsymbol{x}_2$ or $\boldsymbol{x}_3$ as the line-of-sight, therefore we have some 510 velocity dispersion profiles for each of the low-density models and 450 for the SCDM model.

The shape of $\sigma_v(r_p)$ is quantified by the ratios of $\sigma_v(r_p)$ in different bins (cf. den Hartog & Katgert 1994). For convenience, let us denote $\sigma_v(1)$ for $\sigma_v(r_p)$ in the first bin $0 < r_p < 0.2\,h^{-1}$Mpc, $\sigma_v(2)$ for that in the second bin $0.2 < r_p < 0.5\,h^{-1}$Mpc and $\sigma_v(3)$ for that in the third bin $0.5 < r_p < 1.0\,h^{-1}$Mpc. If $\sigma_v(1)/\sigma_v(3) > \sigma_v(1)/\sigma_v(2) > 1$, the velocity dispersion decreases with the projected radius $r_p$ (hereafter Type I); if $\sigma_v(1)/\sigma_v(3) \approx \sigma_v(1)/\sigma_v(2) \approx 1$, the velocity dispersion is nearly a constant (hereafter Type II); if $\sigma_v(1)/\sigma_v(3) < \sigma_v(1)/\sigma_v(2) < 1$, the velocity dispersion increases with $r_p$ (hereafter Type III). There are still other cases in which the shapes of the velocity dispersion profiles are even more complicated than these three cases. The statistical distribution of these ratios of $\sigma_v(r_p)$ can more quantitatively measure the shapes of the dispersion profiles.

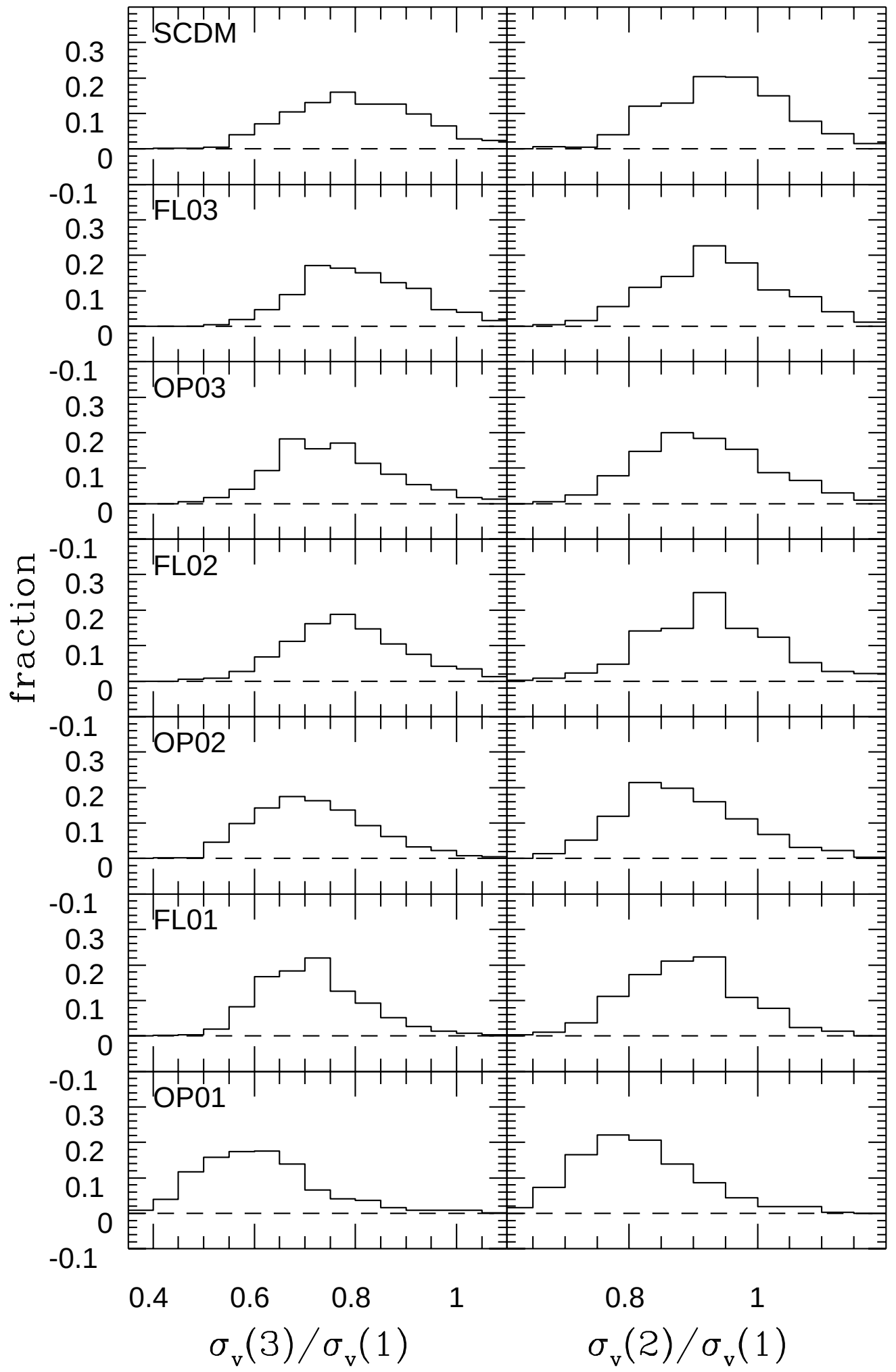

**Figure 1** – The distributions of the ratios $\sigma_v(3)/\sigma_v(1)$ and $\sigma_v(2)/\sigma_v(1)$ in seven cosmological models

**Table 1** The K-S probabilities for the distributions of the dispersion ratios

| | SCDM | FL03 | OP03 | FL02 | OP02 | FL01 | OP01 |
|---|---|---|---|---|---|---|---|
| SCDM | — | 0.16 | $0.64\,10^{-6}$ | $0.57\,10^{-1}$ | $0.12\,10^{-19}$ | $0.11\,10^{-24}$ | 0.00 |
| FL03 | 0.24 | — | $0.20\,10^{-7}$ | $0.30\,10^{-1}$ | $0.60\,10^{-23}$ | $0.18\,10^{-28}$ | 0.00 |
| OP03 | $0.59\,10^{-6}$ | $0.27\,10^{-4}$ | — | $0.20\,10^{-3}$ | $0.21\,10^{-4}$ | $0.76\,10^{-8}$ | 0.00 |
| FL02 | $0.42\,10^{-3}$ | $0.37\,10^{-1}$ | $0.24\,10^{-1}$ | — | $0.44\,10^{-16}$ | $0.15\,10^{-21}$ | 0.00 |
| OP02 | $0.48\,10^{-20}$ | $0.15\,10^{-17}$ | $0.31\,10^{-5}$ | $0.86\,10^{-12}$ | — | 0.48 | $0.39\,10^{-38}$ |
| FL01 | $0.16\,10^{-17}$ | $0.31\,10^{-13}$ | $0.63\,10^{-4}$ | $0.19\,10^{-7}$ | 0.18 | — | $0.51\,10^{-42}$ |
| OP01 | 0.00 | 0.00 | 0.00 | 0.00 | $0.20\,10^{-22}$ | $0.55\,10^{-30}$ | — |

In Figure 1, we present the distributions of the velocity dispersion ratios $\sigma_v(3)/\sigma_v(1)$ and $\sigma_v(2)/\sigma_v(1)$ for the seven models. The average values of $\sigma_v(2)/\sigma_v(1)$ and $\sigma_v(3)/\sigma_v(1)$ are 0.95 and 0.80 (SCDM), 0.95 and 0.81 (FL03), 0.92 and 0.77 (OP03), 0.93 and 0.79 (FL02), 0.89 and 0.72 (OP02), 0.89 and 0.72 (FL01), and 0.81 and 0.61 (OP01) respectively. Therefore, the velocity dispersion profiles in all seven models, on average, are Type I profiles. Furthermore, the profiles in the models of lower $\Omega$ are steeper (i.e. decreases faster with $r_p$) than in the models of higher $\Omega$. For a given $\Omega$, the profiles are flatter in the flat model than in the open model. All these results are consistent with JMBF95 on the cluster density profiles. To assess more quantitatively the differences among the shape distributions of Fig. 1, we have performed a Kolmogorov-Smirnov (K-S) test. The probabilities $P(KS)$ that two distributions draw from the same parent distribution are given in Table 1, with those listed above the diagonal for the $\sigma_v(3)/\sigma_v(1)$ distribution and those below the diagonal for the $\sigma_v(2)/\sigma_v(1)$ distribution. The shape distributions of the open models ($\Omega \leq 0.3$) are different from those of the SCDM model at a very high confidence level ($\geq 1-10^{-6}$). However, the shape distributions of the two flat models of $\Omega \geq 0.3$ are indistinguishable even at the confidence level $\sim$ 90%. The difference between FL02 and SCDM in the shape distributions is much smaller than that between FL02 and OP02. The results indicate that the shapes of the velocity dispersion profiles can more effectively discriminate between models without $\Lambda$ than models with $\Lambda = 1 - \Omega$. These dependences, in principle, can be used to constrain the cosmological parameters $\Omega$ and $\Lambda$ if the velocity dispersion profiles are available for a sample of real clusters. Our results are not difficult to understand, because the density profiles have shown quite similar shapes in flat models of different $\Omega$, and quite different shapes in open models of different $\Omega$ (JMBF95; Crone et al. 1994).

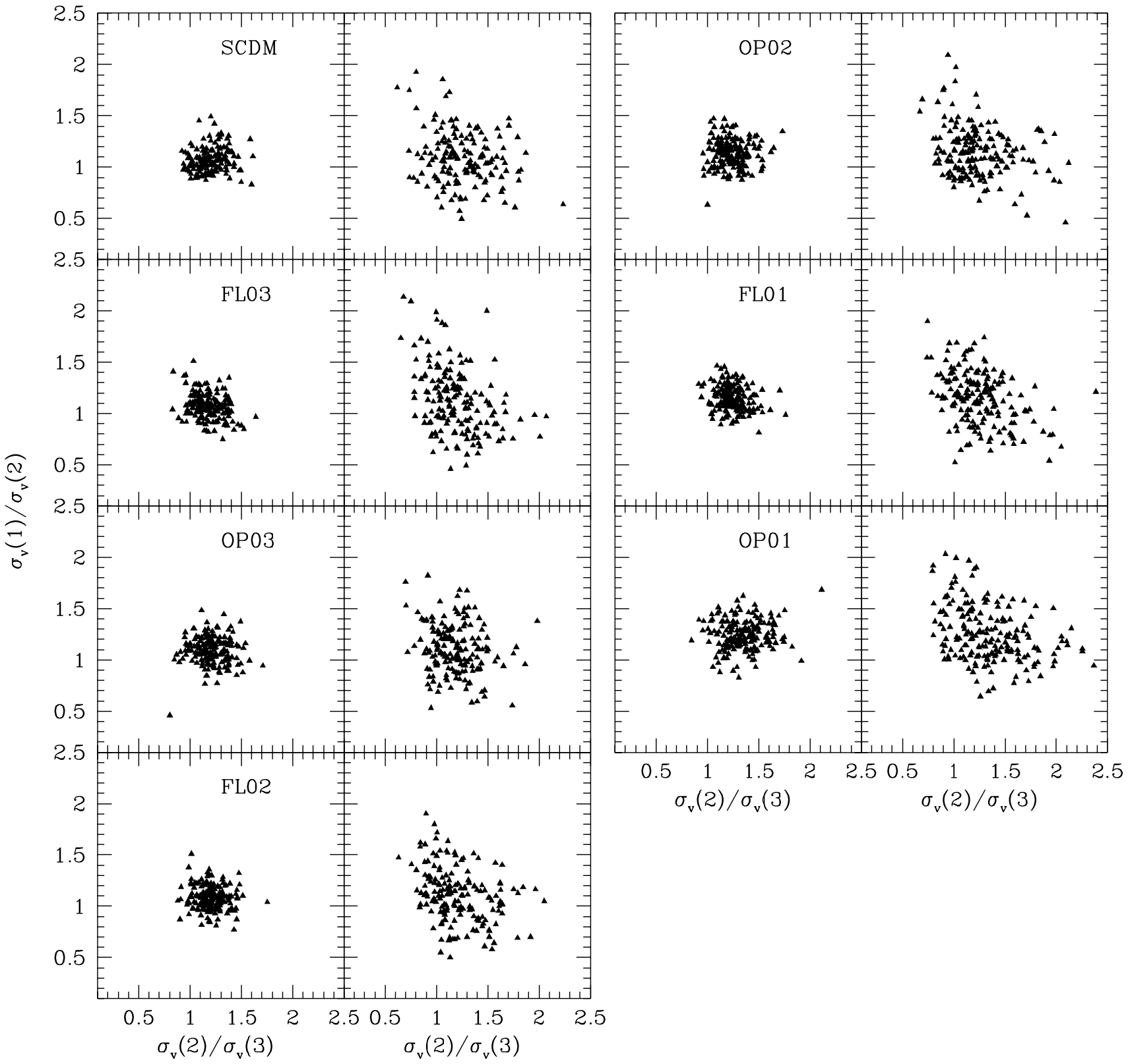

**Figure 2** – The scatter plot of $\sigma_v(1)/\sigma_v(2)$ versus $\sigma_v(2)/\sigma_v(3)$ of the simulated clusters. In the left panels, all cluster members are used to calculate the velocity dispersion profile; while in the right panels, only $\sim$ 100 members are used and an uncertainty of $0.1\,h^{-1}$Mpc is allowed for positions of cluster centers.

However as pointed out in the introduction, the velocity fields around clusters in the real universe cannot be sampled so densely in current observations as in our simulations. To show the importance of the sampling effect, we dilute the model clusters by randomly choosing 100 members† per cluster within projected distance $r_p$ less than $2\,h^{-1}$Mpc. On average, this sampling is already better than current observations. The determination of cluster centers in the observations normally is not as accurate as in our simulations, so we further allow a random displacement for the center positions of the simulated clusters on the projected surface. The displacements obey a Gaussian distribution with zero mean and variance $(0.1\,h^{-1}\mathrm{Mpc})^2$. We have checked the influence of the two effects (sampling and center displacement) separately, and found that the sampling has much more influence on the determination of $\sigma_v(r)$. However, including the center displacement can more accurately simulate the observations. For simplicity, in the following we will not distinguish these two effects and call them together the sampling effect.

The sampling effect is illustrated by Figure 2, where we show the scatter distributions of $\sigma_v(1)/\sigma_v(2)$ versus $\sigma_v(2)/\sigma_v(3)$ before and after the sampling effect is introduced. The scatter distributions with the sampling effect are considerably wider than those without the sampling effect. This means that the dispersion profiles have a wider range of shapes due to the sampling effect. The sampling effect may cause some profiles to decrease or increase with $r_p$ much faster.

This effect is further shown by two examples in Figure 3. The first column shows two clusters in redshift space. Their velocity dispersion profiles are represented as solid squares in the third column. It is easily seen that both clusters have a velocity dispersion profile which weakly decreases with $r_p$. In the second column, these two clusters are plotted after the sampling effect is imposed. The clusters in the second column have appearances different from their partners in the first column. The difference can be easily seen in the velocity dispersion profile. The open squares in the third column show the velocity dispersion $\sigma_v(r_p)$ of the diluted clusters: one shows a strongly increasing profile (for $r_p < 1\,h^{-1}$Mpc)and the other shows a strongly decreasing profile. Should one infer the mass distribution from these profiles (open squares), one would certainly get misleading results. In fact, the two clusters have similar shapes of the mass density profile (fourth column; the slope of both profiles is about $-2.3$). Therefore one should be very cautious when extracting the density profiles from such diluted velocity dispersion profiles. Although these two examples show extreme cases, we find them not unusual: if we take another random dilution of the data set, these two clusters may not show such extreme velocity dispersion profiles, but other clusters will. In fact, we have tried nine additional random dilutions. The distributions of $\sigma_v(1)/\sigma_v(2)$ and $\sigma_v(2)/\sigma_v(3)$ in each random dilution are quite similar to those presented in Fig. 2.

† 100 is the expected number. The actual number culled may differ from 100 under Poisson fluctuations

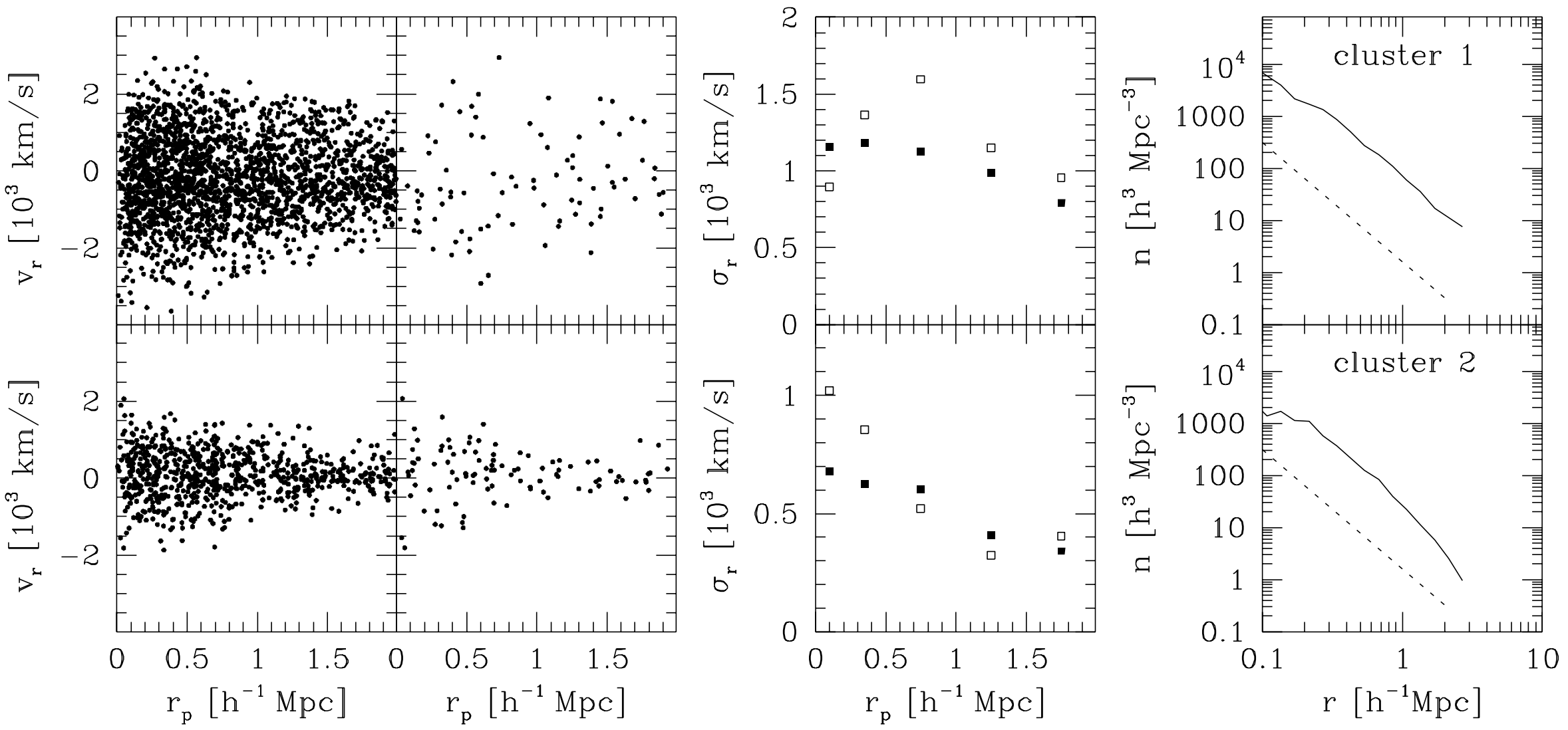

**Figure 3** – An illustration of two clusters. From left, the first column plots cluster members in a diagram of the line-of-sight velocity $v_r$ versus the projected distance $r_p$ from the center; the second column plots the same two clusters but after dilution; the third column shows their velocity dispersion profiles before (solid squares) and after (open squares) the dilution; the last column shows their density profiles $n(r)$ (solid curves), compared with the power-law $n(r) \propto r^{-2.3}$.

**Table 2** The K-S probabilities after random sampling: the mean and $1\sigma$ scatter among 10 samplings

| | SCDM | FL03 | OP03 | FL02 | OP02 | FL01 | OP01 |
|---|---|---|---|---|---|---|---|
| SCDM | — | $0.52 \pm 0.26$ | $0.32 \pm 0.21$ | $0.36 \pm 0.25$ | $0.067 \pm 0.16$ | $0.050 \pm 0.094$ | $(3.3 \pm 7.5)10^{-10}$ |
| FL03 | $0.39 \pm 0.32$ | — | $0.37 \pm 0.24$ | $0.48 \pm 0.36$ | $0.021 \pm 0.035$ | $(8.8 \pm 9.3)10^{-3}$ | $(8.3 \pm 19)10^{-12}$ |
| OP03 | $0.47 \pm 0.31$ | $0.52 \pm 0.28$ | — | $0.58 \pm 0.25$ | $0.14 \pm 0.22$ | $0.11 \pm 0.15$ | $(8.8 \pm 17)10^{-10}$ |
| FL02 | $0.29 \pm 0.20$ | $0.47 \pm 0.26$ | $0.57 \pm 0.20$ | — | $0.089 \pm 0.19$ | $0.15 \pm 0.26$ | $(3.0 \pm 5.3)10^{-11}$ |
| OP02 | $0.19 \pm 0.16$ | $0.16 \pm 0.20$ | $0.20 \pm 0.12$ | $0.21 \pm 0.21$ | — | $0.32 \pm 0.27$ | $(4.4 \pm 12)10^{-5}$ |
| FL01 | $0.14 \pm 0.14$ | $0.20 \pm 0.20$ | $0.21 \pm 0.23$ | $0.29 \pm 0.35$ | $0.41 \pm 0.28$ | — | $(6.4 \pm 17)10^{-5}$ |
| OP01 | $(2.4 \pm 6.2)10^{-3}$ | $(9.2 \pm 20)10^{-4}$ | $(3.0 \pm 7.5)10^{-3}$ | $(4.8 \pm 14)10^{-3}$ | $0.067 \pm 0.11$ | $0.14 \pm 0.23$ | — |

We use the K-S test to quantify how the velocity dispersion profiles are distinguishable among different models when the sampling is introduced. In this test we choose only one direction as the line-of-sight, so we have about 170 profiles for each low-density model and 150 for the SCDM model. Note that this sample is already very large compared with the observational samples recently available. Because the sampling is a random process, statistical fluctuations must exist in the calculated probabilities $P(KS)$. Therefore we have done ten realizations for the random sampling. In Table 2, we present the mean values and the standard $1\sigma$ scatters of $P(KS)$ among the ten realizations. The table shows that the velocity dispersion profiles are not distinguishable among models of $\Omega \geq 0.2$ (90% CL). However, the OP01 model is significantly different from all other six models in the distributions of $\sigma_v(1)/\sigma_v(2)$ and $\sigma_v(1)/\sigma_v(3)$. Therefore we expect that the shape of the velocity dispersion profiles can only effectively discriminate between cosmological models of $\Omega < 0.2$ if the cluster sample consists of $\sim 100$ clusters with only $\sim 100$ redshifts per cluster.

## 4 DISCUSSION AND CONCLUSIONS

In this paper, we have investigated the shapes of the velocity dispersion profiles of clusters in seven CDM cosmological models. On average, the velocity dispersion decreases with $r_p$ in every model. The slopes of the profiles are different in different models. For the models without a cosmological constant, the profiles are steeper in lower-$\Omega$ models than in a higher-$\Omega$ model. Adding a cosmological constant makes the profiles flatter: the profiles are flatter in a low-density flat model than in the corresponding open model of the same $\Omega$. The difference between the SCDM model and a low-density flat model in the shape of the velocity dispersion profiles is much smaller than that between the SCDM and the corresponding open model. These results are consistent with the results found by JMBF95 for the density profiles of clusters. Therefore, the shape of the velocity dispersion profiles, in principle, can be used to test the cosmological models and to constrain the cosmological parameters.

For the time being, a sample of $\sim 100$ clusters with $\sim 100$ redshifts per cluster still represents the largest observational sample available for the analysis of the velocity dispersion profiles. To test the discriminative power of such a sample between cosmological models, we simulated the observation by randomly diluting the clusters to 100 members per cluster. We found that the discriminative power of the velocity dispersion profiles with such a sample is rather limited: the five models with $\Omega \geq 0.2$ are not distinguishable in the profile distributions. Only the OP01 model is still significantly different from the other six models. About 100 members per cluster are generally insufficient to give an unbiased estimate of the velocity dispersion profile. One should be especially cautious if he tries to infer the density distribution and/or the dynamics around clusters from such a diluted velocity dispersion profile.

In this work we have assumed that the cluster members are defined unambiguously. In real observations, however, the cluster members are always difficult to identify due to the projection effects (e.g. Frenk et al. 1990; Bird & Beers 1993). The projection effects can only reduce the discriminative power of the velocity dispersion profiles. We have not examined the possibilities that galaxies are a biased tracer of the underlying dark matter in the coordinate space and/or in the velocity space. With the present knowledge of galaxy formation, we still don't know how galaxies are biased relative to the underlying matter in the phase space distribution. The bias effect would complicate the comparison of the model velocity dispersion profiles with the observational ones. However the sampling effect discussed in the paper should be important regardless of whether galaxies are formed in a biased manner or not.

## ACKNOWLEDGMENTS

YPJ acknowledges the support of an Alexander-von-Humboldt Research Fellowship.